\documentstyle[preprint,aps,prl,epsf]{revtex}
  \def\Eq.#1{Eq.~(\ref{#1})}
  \def\vec#1{\mbox{\boldmath $#1$}}
  \def\gsl#1{\rlap{\slash}#1} 
  \def\tr{{\rm tr}}
  \def\gvec{\vec \gamma}

  \def\w{\omega_0}
  \def\wp{\omega_+}
  \def\wm{\omega_-}
  \def\MBS{{M_B^2}}
  \def\MBQ{{M_B^4}}
  \def\MN{M_N}
  \def\M{M}
  \def\E{E_{\small\vec q}}
  \def\EP{E_{\small\vec q'}}
  \def\MB{M_B}
  \def\vic{\sim}
  \def\V{v}
  \def\I{\,{\rm I}\,}
  \def\u{q_2}
  \def\uu{q_2q_2}
  \def\d{q_1}
  \def\dd{q_1q_1}

\renewcommand{\thanks}{\footnote}
\def\tocite#1{$^{\hbox{\,-}}$\kern-.04em\cite{#1}}

\def\JLone<#1,#2>{#1}
\def\JLtwo<#1,#2,#3>{#2}
\def\JLyear<#1,#2,#3,#4>{#3}
\def\JLpage<#1,#2,#3,#4>{#4}

\def\Jpage<#1,#2,#3>{#3}
\def\andvol#1{{\bf \JLone<#1>} (\JLtwo<#1>), \Jpage<#1>}
\def\PTP#1{Prog.\ Theor.\ Phys.\ \andvol{#1}}

\def\PR#1{Phys.\ Rev.\ \andvol{#1}}
\def\PRL#1{Phys.\ Rev.\ Lett.\ \andvol{#1}}
\def\PL#1{Phys.\ Lett.\ \andvol{#1}}
\def\NP#1{Nucl.\ Phys.\ \andvol{#1}}

\def\ZP#1{Z. Phys.\ \andvol{#1}}

\preprint{KEK-TH-910}
\title{QCD sum rules for hyperon-nucleon interactions}
\author{Yoshihiko Kondo~\hbox{$^\dagger$} and Osamu Morimatsu~\hbox{$^\ddagger$}\\
{\it {$^\dagger$}~Kokugakuin University, Higashi, Shibuya, Tokyo 150-8440, Japan}\\
{\it {$^\ddagger$}~Institute of Particle and Nuclear Studies, High Energy Accelerator Research Organization, Tukuba, Ibaragi 305-0801, Japan}}

\begin{document}
\maketitle

\begin{abstract}
We investigate the hyperon-nucleon interactions in the QCD sum rule starting from the nucleon matrix element of the hyperon correlation function.
Through the dispersion relation, the correlation function in the operator product expansion (OPE) is related with its integral over the physical energy region.
The dispersion integral around the hyperon-nucleon ($YN$) threshold is identified as a measure of the interaction strength in the $YN$ channel.
The Wilson coefficients of the OPE for the hyperon correlation function are calculated.
The obtained sum rules relate $YN$ interaction strengths to the nucleon matrix elements of the quark-gluon composite operators, which include strange quark operators as well as up and down quark operators.
It is found that the $YN$ interaction strengths are smaller than the $NN$ interaction strength since the nucleon matrix elements of strange quark operators are smaller than those of up and down quark operators.
Among $YN$ channels $\Lambda N$ channel has stronger interaction than $\Sigma N$ and $\Xi N$ channels.
Also found is that the interaction strength is greater in the $\Sigma^+N$ ($\Xi^0 N$) channel than in the $\Sigma^-N$ ($\Xi^-N$) channel since the nucleon matrix elements of up quark operators are greater than those of down quark operators.
The spin-dependent part is much smaller than the spin-independent part in the $YN$ and $NN$ channels.
The results of the sum rules are compared with those of the phenomenological meson-exchange models.
\end{abstract}

\newpage
\section{Introduction}
One of the most important goals of studying strong interaction physics is to understand the behaviour of hadrons and hadronic interactions on the basis of the quantum-chromo-dynamics (QCD).
One of the powerful tools for this purpose is the QCD sum rule proposed by Shifman, Vainshtein and Zakharov~\cite{rf:SVZ}.
Using the method of the QCD sum rule one can obtain relations between the properties of hadrons and the vacuum matrix elements of the quark-gluon composite operators~\cite{rf:SVZ,rf:RRY}.

The idea of the QCD sum rule has been extended to the investigation of hadronic interactions~\cite{rf:KM,rf:Koike,rf:KMN,rf:KH,rf:KMII}.
In particular, in Ref.~\cite{rf:KMII} the present authors have formulated the QCD sum rules for spin-dependent nucleon-nucleon interactions and have studied their physical implications.
The basic object of the study in the work has been the nucleon matrix element of the correlation function of the nucleon interpolating field.
In the deep Euclidean region the correlation function is expressed in terms of the nucleon matrix elements of the quark-gluon composite operators using the operator product expansion (OPE), which is related with its integral over the physical region by means of the dispersion relation.
The dispersion integral of the correlation function around the nucleon threshold has been investigated in detail.
It has turned out that the integral can be identified as a measure of the nucleon-nucleon interaction strength, which is proportional to the scattering length in the small scattering length limit and to the one half of the effective range in the large scattering length limit.
The obtained sum rules relate the nucleon-nucleon interaction strengths with the nucleon matrix elements of the quark-gluon composite operators.
The sum rules tell us that the interaction strength in the spin-singlet channel is weaker than in the spin-triplet channel, but that the spin-dependent part of the interaction strength is considerably smaller than the spin-independent part.
Experimentally, it has been found that there is a loosely bound state, deuteron, in the spin-triplet channel, while there is an almost bound state in the spin-singlet channel, which implies that the interaction is slightly stronger in the spin-triplet channel than in the spin-singlet channel.
This is consistent with the result of sum rules.

It is straightforward to apply the method of Ref.~\cite{rf:KMII} to hyperon-nucleon channels just by replacing the nucleon interpolating field by the hyperon interpolating field.
The purpose of the present paper is to carry out this procedure.
Physically, it is extremely important to understand the various hyperon-nucleon and nucleon-nucleon interactions simultaneously in order to achieve a unified view of hadronic interactions.
It is also important in the application to hypernuclear physics since hyperon-nucleon interactions are exprimentally not well known.
The paper is organized as follows.
In Sec. II, we review the formalism in the hyperon-nucleon ($YN$) channel for the sake of completeness.
In Sec. III, we calculate the Wilson coefficients of the OPE for the hyperon correlation function and explain how the nucleon matrix elements of the quark-gluon composite operators are determined.
The obtained sum rules for the hyperon-nucleon interaction strengths are presented and discussed in Sec. IV.
Finally, we summarize the paper in Sec. V.

\section{Phisical Contents of Correlation Functions}

We first consider the spin-dependent hyperon correlation function, $\Pi(q \hat p s)$,
\begin{eqnarray*}
\Pi(q \hat p s)=-i\int d^4x e^{iqx}\langle\hat p s|T(\psi(x)\bar\psi(0))|\hat p s\rangle ,
\end{eqnarray*}
where  $|\hat p s\rangle $ is the one-nucleon state with momentum $\hat p$ and spin $s$ ($\hat p^2=\MN^2$, $s^2=-1$ and $\hat ps=0$, where $\MN$ is the nucleon mass) normalized as  $\langle\hat p s|\hat p' s' \rangle =(2\pi)^3\delta^3(\vec p-\vec p')\delta_{ss'}$ and $\psi$ is the normalized hyperon field operator.
In this paper, momentum with $\;\hat{ }\;$ represents the on-shell momentum.
Later, the normalized nucleon field, $\psi$, is replaced by the unnormalized hyperon interpolating field (quark-gluon composite field), $\eta$.
The following discussion, however, holds as it is for the interpolating field, except for the normalization.
Naively, the dispersion relation for the correlation function, $\Pi(q \hat p s)$, is written as
\begin{eqnarray}\label{eq:dr}
\Pi(q \hat p s)=-{1 \over \pi}\int^\infty_{-\infty}dq'_0{1 \over q_0-q'_0+i\eta}{\rm Im}\Pi(q' \hat p s) ,
\end{eqnarray}
where $q'=(q'_0, \vec q)$.
Throughout this paper, whenever we take the imaginary part of a quantity, we approach the real energy axis from above in the complex energy plane.
Therefore, strictly speaking, ${\rm Im}\Pi$ is the imaginary part of the retarded correlation function.
The QCD sum rules are obtained by evaluating the left-hand side of Eq.~(\ref{eq:dr}) by the OPE and expressing the right-hand side in terms of physical quantities.

Let us consider the singularities of $\Pi(q \hat p s)$ as functions of $q_0$.
In the complex $q_0$ plane, $\Pi(q \hat p s)$ has a branch cut from the lowest $B=2$ continuum threshold to the right and another branch cut starting from the lowest $B=0$ continuum threshold to the left where $B$ denotes the baryon number.
$\Pi(q \hat p s)$ has second-order poles at $q_0=\pm\sqrt{\vec q^2+\M^2}\equiv\pm \E$, where $\M$ denotes the hyperon mass. In addition, the coefficient of the pole at $q_0=\E$ is the $YN$ T-matrices $T_{+}$:
\begin{eqnarray*}
T_{+}(\hat q r\hat p s;\hat q r\hat p s)
&=&\left.(q_0-\E)^2{M \over\E}\bar u(q r)\gamma_0\Pi(q \hat p s)\gamma_0u(q r)\right|_{q_0=\E},
\end{eqnarray*}
where $u(qr)$ is a positive energy solution of the free Dirac equation for the hyperon.

In order to take out the pole contribution from ${\rm Im}\Pi(q \hat p s)$ it is convenient to define off-shell $YN$ T-matrices by
\begin{eqnarray}\label{eq:offT}
&&T_{+}(q' r'\hat p' s';q r\hat p s)\cr
&=&-i\int d^4x e^{iq'x}\sqrt{M \over\EP}\bar u(q' r')(\gsl{q'}-M)\langle\hat p' s'|T(\psi(x)\bar\psi(0))|\hat p s\rangle(\gsl{q}-M)\sqrt{M \over\E}u(q r).
\end{eqnarray}
Note that Eq.~(\ref{eq:offT}) is just a definition of the T-matrix off the mass shell, but the LSZ reduction formula shows rigorously that it is the T-matrix on the mass shell.

In order to eliminate the pole contribution at $q_0=-\E$ from the correlation function, we introduce the projection operators $\Lambda_+$ by
\begin{eqnarray*}
&&\Lambda_+(q r)=u(q r)\bar u(q r)={\gsl{\hat q}+M \over 2M}{1+\gamma_5\gsl{r} \over 2}
\end{eqnarray*}
which have the properties
\begin{eqnarray*}
&&\Lambda_+^2(q r)=\Lambda_+(q r),\cr
&&\gamma_0\Lambda_+(q r)\gamma_0=\Lambda_+(\bar q \bar r).
\end{eqnarray*}
where $\bar q=(q_0,-\vec q)$ and $\bar r=(-r_0,\vec r)$.

Then we define the projected correlation functions by
\begin{eqnarray}\label{pCF}
\Pi_+(q r \hat p s)
&=&{M\over\E}\tr\left\{\Lambda_+(\bar q \bar r)\Pi(q \hat p s)\right\}.
\end{eqnarray}
The projected correlation functions are related to the off-shell T-matrices as
\begin{eqnarray*}
\Pi_+(q r \hat p s)={T_+(q r \hat p s) \over (q_0-\E)^2} ,
\end{eqnarray*}
where $T_+(q r \hat p s)\equiv T_+(q r \hat p s;q r \hat p s)$.
Clearly, $\Pi_+(q r \hat p s)$ has a second-order pole at $q_0=\E$ but not at $q_0=-\E$.

We next consider the dispersion relation for ${q_0-\E \over q_0}\Pi_+(qr \hat ps)$ ,
\begin{eqnarray}\label{eq:PiII}
&&{q_0-\E \over q_0}\Pi_+(qr \hat ps)=-{1 \over \pi}\int^\infty_{-\infty}dq'_0{1 \over q_0-q_0'+i\eta}{\rm Im}\left\{{q'_0-\E \over q'_0}\Pi_+(q'r \hat ps)\right\}.\qquad
\end{eqnarray}
Symmetrizing Eq.~(\ref{eq:PiII}) we obtain
\begin{eqnarray}\label{eq:QSR}
 {q_0-\E \over 2q_0}\Pi_+(qr \hat ps)
 +(q_0\rightarrow-q_0)
={1 \over \pi}\int^\infty_{-\infty}dq'_0{1 \over (q_0+i\eta)^2-q'^2_0}
(q'_0-\E){\rm Im}\Pi_+(q'r \hat ps),
\end{eqnarray}
where
\begin{eqnarray}\label{eq:ImPi}
(q_0-\E){\rm Im}\Pi_+(q r \hat p s)
=-\pi\delta(q_0 - \E)\left.{\rm Re}T_+(q r \hat p s)\right|_{q_0=\E}+
{{\rm P} \over q_0 - \E}{\rm Im}T_+(q r \hat p s).
\end{eqnarray}
Applying the Borel transformation,
\begin{eqnarray*}
       L_B\equiv
       \lim_{{n\rightarrow\infty \atop -q_0^2\rightarrow\infty}
       \atop -q_0^2/n = \MBS}
       {(q_0^2)^n\over(n-1)!}\left(-{d\over dq_0^2}\right)^n ,
\end{eqnarray*}
to both sides of Eq.~(\ref{eq:QSR}), we obtain
\begin{eqnarray}\label{BSR}
& &L_B\Big[{q_0-\E \over 2q_0}\Pi_+(qr \hat ps)
          +(q_0\rightarrow-q_0)
      \Big]\cr
&=&-{1 \over \pi}\int^\infty_{-\infty}dq'_0
    {1\over\MBS}\exp\left(-{q_0'^2\over\MBS}\right)
    (q'_0-\E){\rm Im}\Pi_+(q' r \hat p s) ,
\end{eqnarray}
where $\MB$ is the Borel mass.
In order to derive the Borel sum rules we must evaluate the left-hand side by the OPE and parameterize the right-hand side in terms of physical quantities.

We study the behaviour of the two terms on the left-hand side in \Eq.{eq:ImPi}.
For this purpose it is important to note that the off-shell optical theorem holds for $T$.
When the center-of-mass energy is above the threshold of the $YN$ channel and below the threshold of the next channel, only the $YN$ states contribute in the intermediate states, and the off-shell optical theorem is simplified as
\begin{eqnarray}\label{eq:OT}
{\rm Im}T_+(q \hat p;q \hat p)
=-\pi\int {d^3p_n\over(2\pi)^3}{d^3q_n\over(2\pi)^3}(2\pi)^3
\delta^4(\hat p+q-\hat p_n-\hat q_n)
T_+(q \hat p;\hat q_n\hat p_n)
T_+(\hat q_n\hat p_n;q \hat p).\cr
\end{eqnarray}
In order to simplify the notation we introduce the scattering amplitude $f$ by
\begin{eqnarray*}
f(q'p';qp) = -{\mu'^{1/2}\mu^{1/2} \over 2\pi}T_+(q'p';qp) ,
\end{eqnarray*}
where $\mu={q_0p_0\over q_0+p_0}$ and $\mu'={q'_0p'_0\over q'_0+p'_0}$.
Moreover, we  go to the center-of-mass frame ($\vec q + \vec p = \vec q' + \vec p' =0$) and restrict ourselves to the $s$-wave.
We define the off-shell scattering amplitudes as
\begin{eqnarray*}
f(k) = f(q'\hat p';q \hat p) ,
\end{eqnarray*}
where $|\vec p|=|\vec q|=|\vec p'|=|\vec q'|=0$,
$p_0=p'_0=\MN$, $q_0=q'_0=\sqrt{\M^2+k^2}+\sqrt{\MN^2+k^2}-\MN$.
From Eq.~(\ref{eq:OT}) we have
\begin{eqnarray*}\label{eq:Imf}
&&{\rm Im}{1 \over f(k)} = -k + bk^3 + O(k^5),
\end{eqnarray*}
and find that the off-shell scattering amplitude, $f(k)$, has the form
\begin{eqnarray*}\label{eq:fii}
f(k) = {1 \over i\left\{-k+bk^3+O(k^5)\right\}+
\left\{{1 \over a} + {1 \over 2}\tilde r k^2 + O(k^4)\right\}},
\end{eqnarray*}
where $a$ is the scattering length.
It should be noted that $\tilde r$ is different from the effective range $r$,
but $\tilde r$ coincides with $r$ in the limit $a \rightarrow \infty$:
\begin{eqnarray*}
\tilde r = r + O\left({1 \over a}\right).
\end{eqnarray*}

We proceed to the integral of the right-hand side of Eq.~(\ref{BSR}), $I$, in the vicinity of $q_0 =\M$,
\begin{eqnarray*}
I&=&-{1 \over \pi}\int_{\vic\M}dq'_0
    {1\over\MBS}\exp\left(-{q_0'^2\over\MBS}\right)
    (q'_0-\M){\rm Im}\Pi_+.
\end{eqnarray*}
The integral, $I$, can be decomposed as
\begin{eqnarray}\label{eq:III}
 I=I_{t}+I_{c}\ (+I_{b}).
\end{eqnarray}
In Eq.~(\ref{eq:III}), the first term, $I_{t}$, is the threshold contribution, given by
\begin{eqnarray*}
I_{t}=-{1\over\MBS}\exp\left(-{\M^2\over\MBS}\right)2\pi{\M+\MN\over\M\MN}a.
\end{eqnarray*}
The second term, $I_{c}$, is the continuum contribution, given by
\begin{eqnarray*}
I_{c}=-{1 \over \pi}\int_{\vic\M} dq'_0{1\over\MBS}\exp\left(-{q_0'^2\over\MBS}\right)
{{\rm P} \over q'_0-\M}\left\{-{2\pi}{q'_0+\MN\over q'_0\MN}{\rm Im}f^{cut}_2\right\},
\end{eqnarray*}
where
\begin{eqnarray*}
{\rm Im}f^{cut}_2 = {k\over {1 \over a^2} + \left(1+{\tilde r\over a}+{b\over a^2}\right)k^2+O(k^4)}\theta(q_0-M).
\end{eqnarray*}
The last term, $I_{b}$, is the bound-state contribution, which has to be taken into account if there is a bound state, given by
\begin{eqnarray*}
I_{b}=-{1 \over \pi}\int_{\vic\M} dq'_0{1\over\MBS}\exp\left(-{q_0'^2\over\MBS}\right)
{{\rm P} \over q'_0-\M}\left\{-{2\pi}{q'_0+\MN\over q'_0\MN}{\rm Im}f^{pole}_2\right\},
\end{eqnarray*}
where
\begin{eqnarray*}
{\rm Im}f^{pole}_2 
 &=& -i\pi\left\{\left.{\partial\over\partial k}\left({1 \over f_2}\right)\right|_{k=i\kappa_0}\right\}^{-1}\delta(\kappa-\kappa_0)\cr
 &\equiv& \pi c \delta(\kappa-\kappa_0) ,
\end{eqnarray*}
where $\kappa=ik$ and $i\kappa_0$ is the pole momentum, $1/f_2(i\kappa_0)=0$.

By performing the integral, the continuum contribution becomes
\begin{eqnarray*}
I_c^{YN}
&\approx&
{1\over\MBS}\exp\left(-{\M^2\over\MBS}\right){\M+\MN\over\M\MN}
{2\pi|a|\over\sqrt{
1+{\tilde r \over a}+{b\over a^2}+{1\over 2a^2\M^2}+\left(1+{\M\over\MN}\right){1\over a^2\MBS}}},
\end{eqnarray*}
which is simplified in two limits of $a$ as
\begin{eqnarray*}
I_c^{YN}\rightarrow\left\{
\begin{array}{ll}
{1\over\MBS}\exp\left(-{\M^2\over\MBS}\right)2\pi{\M+\MN\over\M\MN}
{a^2\over\sqrt{{1\over2\M^2}+b+\left(1+{\M\over\MN}\right){1\over\MBS}}}
   &\quad(a \rightarrow 0),\cr
{1\over\MBS}\exp\left(-{\M^2\over\MBS}\right)2\pi{\M+\MN\over\M\MN}
|a|\left(1-{r\over2a}\right)
   &\quad(a \rightarrow \infty),
\end{array}
\right.
\end{eqnarray*}

Similarly, the bound-state contribution becomes
\begin{eqnarray*}
I_b
&=&{1 \over \pi}
{\kappa_0(\sqrt{\M^2-\kappa_0^2}+\sqrt{\MN^2-\kappa_0^2})\over\sqrt{\M^2-\kappa_0^2}\sqrt{\MN^2-\kappa_0^2}},
{1\over\MBS}\exp\left(-{\omega'^2\over\MBS}\right)
{1 \over \omega' - \M}2\pi^2{\omega'+\MN\over\omega'\MN}c,
\end{eqnarray*}
where $\omega'=\sqrt{\M^2-\kappa_0^2}+\sqrt{\MN^2-\kappa_0^2}-\MN$
In the limit $a\rightarrow\infty$,
\begin{eqnarray*}
&&\kappa_0 \rightarrow -{1 \over a}\left[1-{r\over2a}+O\left({1 \over a^2}\right)\right],\cr
&&c \rightarrow \left(1-{r \over a}\right)+O\left({1 \over a^2}\right),
\end{eqnarray*}
and the bound-state contribution is simplified as
\begin{eqnarray*}
I_b\rightarrow
{1\over\MBS}\exp\left(-{\M^2\over\MBS}\right)2\pi{\M+\MN\over\M\MN}2a\left(1-{r \over 2a}\right)+O\left({1 \over a}\right).
\qquad (a \rightarrow \infty)
\end{eqnarray*}

Let us suppose that one can freely change the interaction strength of hyperon-nucleon and examine how the integral $I$ should change as a function the interaction strength.
When the interaction is weak, the scattering length is also small, and the integral $I$ is dominated by $I_{t}$:
\begin{eqnarray*}
I&=&I_{t}+I_{c}\cr
 &=&-{1\over\MBS}\exp\left(-{\M^2\over\MBS}\right)2\pi{\M+\MN\over\M\MN}a+O(a^2).
\end{eqnarray*}
As the interaction becomes stronger, the scattering length increases and the integral, $I$, also increases.
As the interaction strength increases further, the scattering length eventually diverges when the bound state is just formed.
Just before the bound state is formed, the integral $I$ becomes
\begin{eqnarray*}
I&=&I_{t}+I_{c}\cr
&=&
-{1\over\MBS}\exp\left(-{\M^2\over\MBS}\right)2\pi{\M+\MN\over\M\MN}{r\over2}+O\left({1 \over a}\right),
\end{eqnarray*}
and just after the bound state is formed, it becomes
\begin{eqnarray*}
I&=&I_{t}+I_{c}+I_{b}\cr
&=&
-{1\over\MBS}\exp\left(-{\M^2\over\MBS}\right)2\pi{\M+\MN\over\M\MN}{r\over2}+O\left({1 \over a}\right).
\end{eqnarray*}
This shows that before and after the bound state is formed the integral is continuous, though the scattering length diverges with opposite signs.
This observation leads us to conjecture that the integral around the threshold is a measure of the $YN$ interaction strength.

Here we study the integral in a solvable case, namely the separable potential model:
\begin{eqnarray*}
V={\alpha\over \M}{1\over k^2+\beta^2}{1\over k'^2+\beta^2}.
\end{eqnarray*}
We obtain
\begin{eqnarray*}
I\approx-{1 \over \MBS}\exp\left(-{\M^2\over\MBS}\right){\alpha\over \M\beta^4}+O\left({1\over \MB^4}\right).
\end{eqnarray*}
We find that the integral $I$ is proportional to the potential strength $\alpha$ if the range parameter $\beta$ is fixed.

Based on this conjecture we define the $YN$ interaction strength, $\V$, by
\begin{eqnarray}\label{eq:I}
I&\equiv&-{1\over\MBS}\exp\left(-{\M^2\over\MBS}\right)\V.
\end{eqnarray}

In the dispersion integral, the imaginary part of the correlation function,  ${\rm Im}\Pi_+$, contains the contribution from all possible intermediate states such as those of the $YN$, $YN\pi$ channels and so on.
However, only the $YN$ channel contributes around the threshold.
We assume that the contribution from the $YN$ state is taken into account by the form of the right-hand side of Eq.~(\ref{eq:I}) and that the rest is approximated by an asymptotic form of the correlation function starting from an effective threshold, $\omega_+$,  for the $B=2$ channels other than the $YN$ channel and $\omega_-$ for the $B=0$ channels:
\begin{eqnarray}\label{ImPi}
(q_0-\M){\rm Im}\Pi_+
&=&\lambda^2\pi\delta(q_0-\M)\V \cr
& &+\left\{\theta(-q_0-\omega_-)+\theta(q_0-\omega_+)\right\}
(q_0-\M){\rm Im}{\Pi}_+^{OPE},
\end{eqnarray}
where $\Pi^{OPE}_{+}$ is the asymptotic form of the correlation function in the OPE and the normalization constant $\lambda$ is explicitly included ($\langle 0|\eta(0)|\hat q s\rangle=\lambda u(qs)$). 
This is possible now because the contribution from states other than those of the $YN$ channel is exponentially suppressed compared to the $YN$ contribution. 

\section{OPE of Correlation Functions}

Let us turn to the OPE.
We take the interpolating fields of hyperons~\cite{rf:Ioffe} as
\begin{eqnarray*}
\eta&=&\sqrt{2\over3}\epsilon_{abc}\{
 (u^a C\gamma_\mu s^b)\gamma_5\gamma_\mu d^c
-(d^a C\gamma_\mu s^b)\gamma_5\gamma_\mu u^c\}
\end{eqnarray*}
for the $\Lambda$ and 
\begin{eqnarray*}
    \eta&=&\epsilon_{abc}(\d^a C\gamma_\mu\d^b)\gamma_5\gamma_\mu\u^c
\end{eqnarray*}
for the other hyperons where  
$\d=u$, $\u=s$ for the $\Sigma^+$, 
$\d=d$, $\u=s$ for the $\Sigma^-$, 
$\d=s$, $\u=u$ for the $\Xi^0$ and 
$\d=s$, $\u=d$ for the $\Xi^-$.
Since the interction strength for the $\Sigma^0N$ channel is obtained by averaging those for the $\Sigma^+N$ and $\Sigma^-N$, $\eta$ for the $\Sigma^0$ is not needed here.
$C$ denotes the charge conjugation operator and $a$, $b$ and $c$ are color indices.
We take into account all the operators of dimension less than or equal to four.
We also include four-quark operators of dimension six, since four-quark operators are known to give the largest contribution among higher order operators in the QCD sum rule for the nucleon mass~\cite{rf:RRY,rf:HL}.
The OPE of the correlation functions are given as
\begin{eqnarray}\label{OPE_L}
  & & \Pi^{OPE}(q)\cr
  &=&{1\over4\pi^4}\gamma^\mu\Big[q_\mu q^4\ln(-q^2){1\over16}\langle\I\rangle+
    q^2\ln(-q^2)\pi^2\Big\{-{11\over18}\langle\bar d\gamma_\mu d\rangle
                           -{11\over18}\langle\bar u\gamma_\mu u\rangle
                           -{13\over9}\langle\bar s\gamma_\mu s\rangle
                     \Big\}\cr
  & &\qquad
   +q_\mu q^\nu\ln(-q^2)\pi^2\Big\{-{5\over9}\langle\bar d\gamma_\nu d\rangle
                                   -{5\over9}\langle\bar u\gamma_\nu u\rangle
                                   -{2\over9}\langle\bar s\gamma_\nu s\rangle
                             \Big\}\cr
  & &\qquad
   +q_\mu\ln(-q^2)\pi^2\Big\{
    {1\over8}\left\langle{\alpha_s\over\pi}G^{\alpha\beta}G_{\alpha\beta}
             \right\rangle    
   +{1\over3}(3m_d+m_u-2m_s)\langle\bar dd\rangle \cr
  & &\qquad\qquad
    +{1\over3}(3m_u+m_d-2m_s)\langle\bar uu\rangle
    +{1\over3}(3m_s-2m_d-2m_u)\langle\bar ss\rangle
                       \Big\}\cr
  & &\qquad
   +q^\nu\ln(-q^2)\pi^2\Big\{
    -{1\over6}\left\langle{\alpha_s\over\pi}{\cal S}
                  [G^{\rho}_{\mu}G_{\rho\nu}]\right\rangle \cr
  & &\qquad\qquad
             +{16\over9}i\langle{\cal S}[\bar d\gamma_\mu D_\nu d]\rangle
             +{16\over9}i\langle{\cal S}[\bar u\gamma_\mu D_\nu u]\rangle
             +{28\over9}i\langle{\cal S}[\bar s\gamma_\mu D_\nu s]\rangle
                       \Big\}\cr
  & &\qquad
   +q_\mu{1\over q^2}\pi^4\Big\{ {16\over9}\langle\bar dd\bar ss\rangle
                                +{16\over9}\langle\bar uu\bar ss\rangle
                                -{8\over9}\langle\bar dd\bar uu\rangle
                          \Big\} \Big] \cr
  &+&{1\over4\pi^4}\Big[ q^4\ln(-q^2)\Big\{{1\over24}(2m_u+2m_d-m_s)\langle\I\rangle\Big\}+
      q^2\ln(-q^2)\pi^2\Big\{-{2\over3}\langle\bar dd\rangle
                             -{2\over3}\langle\bar uu\rangle
                             +{1\over3}\langle\bar ss\rangle
                       \Big\}\cr
  & &\qquad  
   +q^\mu\ln(-q^2)\pi^2\Big\{
     {1\over3}(4m_d-2m_u+m_s)\langle\bar d\gamma_\mu d\rangle
    +{1\over3}(4m_u-2m_d+m_s)\langle\bar u\gamma_\mu u\rangle \cr
  & &\qquad\qquad
    -{2\over3}(m_s+m_d+m_u)\langle\bar s\gamma_\mu s\rangle
                       \Big\}\cr
  & &\qquad  
   +q^\mu{1\over q^2}\pi^4\Big\{ 
     {16\over9}(\langle\bar dd\bar u\gamma_\mu u\rangle
                +\langle\bar dd\bar s\gamma_\mu s\rangle)
    +{16\over9}(\langle\bar uu\bar d\gamma_\mu d\rangle
                +\langle\bar uu\bar s\gamma_\mu s\rangle)\cr
  & &\qquad\qquad -{8\over9}(\langle\bar ss\bar d\gamma_\mu d\rangle
                +\langle\bar ss\bar u\gamma_\mu u\rangle)
                       \Big\} \cr
  & &\qquad
   +{1\over q^2}\pi^4\Big\{
     {4\over9}m_u(-2\langle\bar uu\bar dd\rangle
     +12\langle\bar dd\bar ss\rangle+\langle\bar uu\bar ss\rangle)
  \cr& &\qquad\qquad+{4\over9}m_d(-2\langle\bar uu\bar dd\rangle
     +12\langle\bar uu\bar ss\rangle+\langle\bar dd\bar ss\rangle)\cr
  & &\qquad\qquad    +{8\over9}m_s(-\langle\bar dd\bar ss\rangle
     +6\langle\bar uu\bar dd\rangle-\langle\bar uu\bar ss\rangle)
  \cr& &\qquad\qquad+{1\over18}(2\langle\bar uu\rangle+2\langle\bar dd\rangle
        -\langle\bar ss\rangle)
      \left\langle{\alpha_s\over\pi}G^{\alpha\beta}G_{\alpha\beta}\right\rangle
    \Big\} \Big] \cr
  &+&{1\over4\pi^4}\gamma^\mu\gamma_5\Big[ 
    q^2\ln(-q^2)\pi^2\Big\{{1\over18}\langle\bar d\gamma_\mu\gamma_5 d\rangle
                          +{1\over18}\langle\bar u\gamma_\mu\gamma_5 u\rangle
                          +{11\over9}\langle\bar s\gamma_\mu\gamma_5 s\rangle
                     \Big\}\cr
  & &\qquad
   +q_\mu q^\nu\ln(-q^2)\pi^2\Big\{
          -{5\over9}\langle\bar d\gamma_\nu\gamma_5 d\rangle
          -{5\over9}\langle\bar u\gamma_\nu\gamma_5 u\rangle
          -{2\over9}\langle\bar s\gamma_\nu\gamma_5 s\rangle
                             \Big\}\cr
  & &\qquad
   +q^\nu\ln(-q^2)\pi^2\Big\{
      {4\over9}\langle{\cal S}[\bar d\gamma_\mu\gamma_5 iD_\nu d]\rangle
     +{4\over9}\langle{\cal S}[\bar u\gamma_\mu\gamma_5 iD_\nu u]\rangle
     -{20\over9}\langle{\cal S}[\bar s\gamma_\mu\gamma_5 iD_\nu s]\rangle\cr
  & &\qquad\qquad 
  -{1\over3}(m_d+m_u-2m_s)
     \left(\langle\bar d\gamma_5i\sigma_{\mu\nu}d\rangle
    +\langle\bar u\gamma_5i\sigma_{\mu\nu}u\rangle
   -2\langle\bar s\gamma_5i\sigma_{\mu\nu}s\rangle\right)
                       \Big\}\cr
  & &\qquad
   +q^\nu{1\over q^2}\pi^4\Big\{
     -{16\over9}(\langle\bar ss\bar d\gamma_5i\sigma_{\mu\nu}d\rangle
                 +\langle\bar ss\bar u\gamma_5i\sigma_{\mu\nu}u\rangle)
     -{16\over9}(\langle\bar dd\bar s\gamma_5i\sigma_{\mu\nu}s\rangle
                 +\langle\bar uu\bar s\gamma_5i\sigma_{\mu\nu}s\rangle)\cr
  & &\qquad\qquad +{8\over9}
       (\langle\bar dd\bar u\gamma_5i\sigma_{\mu\nu}u\rangle
       +\langle\bar uu\bar d\gamma_5i\sigma_{\mu\nu}d\rangle)
                          \Big\} \Big] \cr
  &+&{1\over4\pi^4}i\gamma_5\sigma^{\mu\nu}\Big[
    q^2\ln(-q^2)\pi^2\Big\{
         {1\over9}\langle\bar d\gamma_5i\sigma_{\mu\nu}d\rangle
        +{1\over9}\langle\bar u\gamma_5i\sigma_{\mu\nu}u\rangle
        -{1\over18}\langle\bar s\gamma_5i\sigma_{\mu\nu}s\rangle
                     \Big\}\cr
  & &\qquad
   +q_\mu q^\rho\ln(-q^2)\pi^2\Big\{
         {4\over9}\langle\bar d\gamma_5i\sigma_{\nu\rho}d\rangle
        +{4\over9}\langle\bar u\gamma_5i\sigma_{\nu\rho}u\rangle
        -{2\over9}\langle\bar s\gamma_5i\sigma_{\nu\rho}s\rangle
                              \Big\}\cr
  & &\qquad
   +q_\mu \ln(-q^2)\pi^2\Big\{
         {1\over9}(3m_s-6m_u-4m_d)\langle\bar d\gamma_\nu\gamma_5 d\rangle
        +{1\over9}(3m_s-6m_d-4m_u)\langle\bar u\gamma_\nu\gamma_5 u\rangle\cr
  & &\qquad\qquad
        -{2\over9}(3m_d+3m_u-m_s)\langle\bar s\gamma_\nu\gamma_5 s\rangle
                              \Big\}\cr
  & &\qquad
   +q_\mu{1\over q^2}\pi^4\Big\{
      {16\over9}(\langle\bar dd\bar u\gamma_\nu\gamma_5 u\rangle
                 +\langle\bar dd\bar s\gamma_\nu\gamma_5 s\rangle)
     +{16\over9}(\langle\bar uu\bar d\gamma_\nu\gamma_5 d\rangle
                 +\langle\bar uu\bar s\gamma_\nu\gamma_5 s\rangle)\cr
  & &\qquad\qquad -{8\over9}(\langle\bar ss\bar d\gamma_\nu\gamma_5 d\rangle
                             +\langle\bar ss\bar u\gamma_\nu\gamma_5 u\rangle)
                          \Big\} \Big]
\end{eqnarray}
for $\Lambda$ and
\begin{eqnarray}\label{OPE}
  & & \Pi^{OPE}(q)\cr
  &=&{1\over4\pi^4}\gamma^\mu\Big[q_\mu q^4\ln(-q^2){1\over16}\langle\I\rangle+ 
    q^2\ln(-q^2)\pi^2\Big\{-{7\over3}\langle\bar\d\gamma_\mu\d\rangle
                           -{1\over3}\langle\bar\u\gamma_\mu\u\rangle\Big\}\cr
  & &\qquad
   +q_\mu q^\nu\ln(-q^2)\pi^2\Big\{-{2\over3}\langle\bar\d\gamma_\nu\d\rangle
                           -{2\over3}\langle\bar\u\gamma_\nu\u\rangle\Big\}\cr
  & &\qquad
   +q_\mu\ln(-q^2)\pi^2\Big\{ m_{\u}\langle\bar\uu\rangle
    +{1\over8}\left\langle{\alpha_s\over\pi}G^{\alpha\beta}G_{\alpha\beta}
             \right\rangle\Big\}\cr
  & &\qquad
   +q^\nu\ln(-q^2)\pi^2\Big\{
    -{1\over6}\left\langle{\alpha_s\over\pi}{\cal S}
                  [G^{\rho}_{\mu}G_{\rho\nu}]\right\rangle 
    +{16\over3}i\langle{\cal S}[\bar\d\gamma_\mu D_\nu\d]\rangle
    +{4\over3}i\langle{\cal S}[\bar\u\gamma_\mu D_\nu\u]\rangle\Big\}\cr
  & &\qquad
   +q_\mu{1\over q^2}\pi^4
    \Big\{{8\over3}\langle\bar\dd 
                          \bar\dd\rangle \Big\} \Big] \cr
  &+&{1\over4\pi^4}\Big[ q^4\ln(-q^2)\Big\{{1\over8}m_{\u}\langle\I\rangle\Big\}+
    q^2\ln(-q^2)\pi^2\{-\langle\bar\uu\rangle\}\cr
  & &\qquad   
   +q^\mu\ln(-q^2)\pi^2\{
    2m_{\u}(\langle\bar\u\gamma_\mu\u\rangle
    -\langle\bar\d\gamma_\mu\d\rangle)\}
   +q^\mu{1\over q^2}\pi^4\Big\{{16\over3}
      \langle\bar\uu 
             \bar\d\gamma_\mu\d\rangle\Big\} \cr
  & &\qquad   
   +{1\over q^2}\pi^4\Big\{8m_{\d}\langle\bar\dd\bar\uu\rangle
   +{16\over3}m_{\u}\langle\bar\dd\bar\dd\rangle
   +{1\over6}\langle\bar\uu\rangle
    \left\langle{\alpha_s\over\pi}G^{\alpha\beta}G_{\alpha\beta}
             \right\rangle
    \Big\}
 \Big] \cr
  &+&{1\over4\pi^4}\gamma^\mu\gamma_5\Big[ 
    q^2\ln(-q^2)\pi^2\Big\{{5\over3}\langle\bar\d\gamma_\mu\gamma_5\d\rangle
                  -{1\over3}\langle\bar\u\gamma_\mu\gamma_5\u\rangle\Big\}\cr
  & &\qquad
   +q_\mu q^\nu\ln(-q^2)\pi^2\Big\{
          -{2\over3}\langle\bar\d\gamma_\nu\gamma_5\d\rangle
          -{2\over3}\langle\bar\u\gamma_\nu\gamma_5\u\rangle\Big\}\cr
  & &\qquad
   +q^\nu\ln(-q^2)\pi^2\Big\{
          -{8\over3}\langle{\cal S}[\bar\d\gamma_\mu\gamma_5 iD_\nu\d]\rangle
          +{4\over3}\langle{\cal S}[\bar\u\gamma_\mu\gamma_5 iD_\nu\u]\rangle\Big\}\cr
  & &\qquad
   +q^\nu{1\over q^2}\pi^4\Big\{
           -{16\over3}\langle\bar\dd 
             \bar\d\gamma_5i\sigma_{\mu\nu}\d\rangle\Big\}
 \Big] \cr
  &+&{1\over4\pi^4}i\gamma_5\sigma^{\mu\nu}\Big[
    q^2\ln(-q^2)\pi^2\Big\{
   {1\over6}\langle\bar\u\gamma_5i\sigma_{\mu\nu}\u\rangle \Big\}
   +q_\mu q^\rho\ln(-q^2)\pi^2\Big\{
   {2\over3}\langle\bar\u\gamma_5i\sigma_{\nu\rho}\u\rangle \Big\}
 \Big\}\cr
  & &\qquad
   +q_\mu\ln(-q^2)\pi^2\Big\{-{2\over3}m_{\u}(
    \langle\bar\u\gamma_\nu\gamma_5\u\rangle
    +3\langle\bar\d\gamma_\nu\gamma_5\d\rangle)\Big\}
  \cr& &\qquad+q_\mu{1\over q^2}\pi^4\Big\{{16\over3}
    \langle\bar\uu 
           \bar\d\gamma_\nu\gamma_5\d\rangle \Big\} \Big]
\end{eqnarray}
for other hyperons where the matrix elements can be taken with any state.
The vacuum-to-vacuum and nucleon-to-nucleon correlation functions, $\Pi_0^{OPE}(q)$ and $\Pi^{OPE}(q \hat p s)$, are obtained by replacing the matrix element of ${\cal O}$, $\langle{\cal O}\rangle$, with the vacuum and nucleon matrix elements, $\langle{\cal O}\rangle_0\equiv\langle0|{\cal O}|0\rangle$ and $\langle{\cal O}\rangle_N\equiv\langle ps|{\cal O}|ps\rangle-\langle ps|ps\rangle\langle{\cal O}\rangle_0$, respectively.

We now discuss the nucleon matrix elements of the quark-gluon operators.
Dimension-three operators are quark bilinear operators, $\bar q q$, $\bar q\gamma_\mu q$, $\bar q\gamma_\mu\gamma_5 q$ and $\bar q\gamma_5\sigma_{\mu\nu} q$.
The nucleon matrix elements, $\langle\bar q q\rangle_N$ and $\langle\bar q\gamma_\mu q\rangle_N$, are spin-independent and have already been discussed in Ref.\cite{rf:DL},
while $\langle\bar q\gamma_\mu\gamma_5 q\rangle_N$ and $\langle\bar q\gamma_5\sigma_{\mu\nu} q\rangle_N$ are spin-dependent and are written in terms of the axial charge, $\Delta q$, and the tensor charge, $\delta q$, as
\begin{eqnarray*}
    &&\langle\bar q\gamma_\mu\gamma_5 q\rangle_N=\Delta q s_\mu ,\cr
    &&\langle\bar qi\gamma_5\sigma_{\mu\nu}q\rangle_N
    =\delta q(s_\mu \hat p_\nu-s_\nu \hat p_\mu)/\hat p_0 .
\end{eqnarray*}
For the vector and scalar charges, we take,
$\langle u^\dagger u\rangle_p=2$, $\langle d^\dagger d\rangle_p=1$,
$\langle s^\dagger s\rangle_p=0$,
$\langle\bar u u\rangle_p=3.46$, $\langle\bar d d\rangle_p=2.96$,
$\langle\bar s s\rangle_p=0.7$~\cite{rf:GLS}.
For the axial and the tensor charges, we use the values,
$\Delta u = 0.638 \pm 0.054$, $\Delta d = -0.347 \pm 0.046$,
    $\Delta s = -0.109 \pm 0.030$,
  $\delta u=0.839 \pm 0.060$, $\delta d=-0.231 \pm 0.055$,
    $\delta s=-0.046 \pm 0.034$,
obtained by the lattice calculations~\cite{rf:FKOU,rf:ADHK}.

Dimension-four operators are gluon operators, ${\alpha_s\over\pi}G^{\mu\nu}G_{\mu\nu}$, ${\alpha_s\over\pi}{\cal S}[G^{\rho}_{\mu}G_{\rho\nu}]$, and quark operators, ${\cal S}[\bar q\gamma_\mu iD_\nu q]$, $\bar q{\cal S}(\gamma_\mu iD_\nu)\gamma_5 q$.
The nucleon matrix elements of the gluon operators are spin-independent and have already been discussed in Ref.\cite{rf:DL,rf:JCFG}
\begin{eqnarray*}
 && \langle{\alpha_s\over\pi}G_{\mu\nu}G^{\mu\nu}\rangle_p =-738 \;{\rm MeV},\qquad
   \langle{\alpha_s\over\pi}{\cal S}[G_{\mu 0}G^{\mu 0}]\rangle_p =-50 \;{\rm MeV}.
\end{eqnarray*}
The matrix elements of the spin-independent operators, ${\cal S}[\bar q\gamma_\mu iD_\nu q]$, have also been discussed in Ref.\cite{rf:HL} 
\begin{eqnarray*}
 && i\langle{\cal S}[\bar u\gamma_\mu D_\nu u]\rangle_p=222\;{\rm MeV},\qquad
    i\langle{\cal S}[\bar d\gamma_\mu D_\nu d]\rangle_p=95\;{\rm MeV},\cr
 &&   i\langle{\cal S}[\bar s\gamma_\mu D_\nu s]\rangle_p=18\;{\rm MeV}.
\end{eqnarray*}
On the other hand, the matrix elements of the operator, $\bar q{\cal S}(\gamma_\mu iD_\nu)\gamma_5 q$, is spin-dependent and is expressed as
\begin{eqnarray*}
 &&\langle\bar q{\cal S}(\gamma_\mu iD_\nu)\gamma_5 q\rangle_N
   = a_1s_{\{\mu} \hat p_{\nu\}},
\end{eqnarray*}
where $a_1$ is related to the first moment of the longitudinal quark-spin distribution, $g_2$, if operators including the quark mass are neglected, which is given at the tree-level by~\cite{rf:KYU}
\begin{eqnarray*}
 a_1=-2\int_0^1dxxg_2(x).
\end{eqnarray*}
Using this expression we have calculated $a_1$ from $g_2$.
Experimental data for $g_2$ are given in Ref.\cite{rf:g2p} over the range $0.075<x<0.8$ and $1.3<Q^2<10\,(({\rm GeV}/c)^2)$ for the proton and in Ref.\cite{rf:g2n} over the range $0.06<x<0.70$ and $1.0<Q^2<17.0\,(({\rm GeV}/c)^2)$ for the neutron.
The results are as follows:
\begin{eqnarray*}
     a^u_1 = 0.05 \pm 0.04,\qquad
     a^d_1 =-0.08 \pm 0.13.
\end{eqnarray*}
Unfortunately $a^s_1$ is not determined by experimental data up to now.
However, absolute values of nucleon matrix elements of strange quark operators are generally much smaller than those of up and down quark operators.
Therefore, we set $a^s_1=0$.

The dimension-six four-quark operators are $\bar qq\bar qq$, $\bar qq\bar q\gamma_\mu q$, $\bar qq\bar q\gamma_\mu\gamma_5 q$ and $\bar qq\bar q\gamma_5\sigma_{\mu\nu} q$.
The nucleon matrix elements of the four-quark operators are approximated to factorize~\cite{rf:KMII} as
\begin{eqnarray*}
&&\langle\bar qq\bar qq\rangle_N=2\langle\bar qq\rangle_0\langle\bar qq\rangle_N,
\qquad \langle\bar qq\bar q\gamma_\mu q\rangle_N=\langle\bar qq\rangle_0\langle\bar q\gamma_\mu q\rangle_N,\cr
&&\langle\bar qq\bar q\gamma_\mu\gamma_5 q\rangle_N=\langle\bar qq\rangle_0\langle\bar q\gamma_\mu\gamma_5 q\rangle_N, \qquad \langle\bar qq\bar q\gamma_5\sigma_{\mu\nu} q\rangle_N=\langle\bar qq\rangle_0\langle\bar q\gamma_5\sigma_{\mu\nu} q\rangle_N,
\end{eqnarray*}
which can be evaluated from the vacuum and nucleon matrix elements of dimension-three quark operators.

For completeness we also list here the values of vacuum condensates and quark masses
\begin{eqnarray*}
&&\langle\bar uu\rangle_0 =\langle\bar dd\rangle_0 = -(250\;{\rm MeV})^3,\quad
   \langle\bar ss\rangle_0 = -(232\;{\rm MeV})^3,\cr
&&\langle{\alpha_s\over\pi}G^2\rangle_0 = (330\;{\rm MeV})^4,\cr
&& m_u=m_d=0,\qquad m_s=120\;{\rm MeV}.
\end{eqnarray*}

\section{Results}
The OPE of the projected correlation function, $\Pi_+^{OPE}$, for the $\Lambda$ or other hyperons is given by substituting \Eq.{OPE_L} and \Eq.{OPE}, respectively into the right-hand side of \Eq.{pCF}. 
Then, substituting $\Pi_+^{OPE}$ into the left-hand side of \Eq.{BSR}, and \Eq.{ImPi} into the the right-hand side of \Eq.{BSR}, respectively, we obtain the Borel sum rules for hyperon-nucleon interaction strengths.
The sum rule for the spin-independent part is
\begin{eqnarray}\label{SR^indep_L}
&&-{1\over\MBS}\exp\left(-{\M^2\over\MBS}\right)\lambda^2\V^{indep.}\cr
 &=&{1\over4\pi^4}\Big\{
    \left(C_2\M\MB -C_3\MBS\right)\cr
 & &\qquad\times
    \Big[ \Big\{
    -{7\over6}\langle\bar d\gamma_0 d\rangle_N
    -{7\over6}\langle\bar u\gamma_0 u\rangle_N
    -{5\over3}\langle\bar s\gamma_0 s\rangle_N
    -{2\over3}\langle\bar dd\rangle_N-{2\over3}\langle\bar uu\rangle_N
    +{1\over3}\langle\bar ss\rangle_N
                         \Big\}\Big]\cr
 & &\qquad
    +\left(C_1\M-C_2\MB\right)
     \Big[\pi^2\Big\{
    {1\over8}\left\langle{\alpha_s\over\pi}G^{\alpha\beta}G_{\alpha\beta}
             \right\rangle_N
    -{1\over6}\left\langle{\alpha_s\over\pi}{\cal S}
                  [G^{\rho}_{0}G_{\rho0}]\right\rangle_N \cr
  & &\qquad\qquad
    +{16\over9}i\langle{\cal S}[\bar d\gamma_0 D_0 d]\rangle_N
    +{16\over9}i\langle{\cal S}[\bar u\gamma_0 D_0 u]\rangle_N
    +{28\over9}i\langle{\cal S}[\bar s\gamma_0 D_0 s]\rangle_N \cr
  & &\qquad\qquad
    +{1\over3}(3m_d+m_u-2m_s)\langle\bar dd\rangle_N
    +{1\over3}(3m_u+m_d-2m_s)\langle\bar uu\rangle_N
 \cr  & &\qquad\qquad
    +{1\over3}(3m_s-2m_d-2m_u)\langle\bar ss\rangle_N
    +{1\over3}(4m_d-2m_u+m_s)\langle\bar d\gamma_0 d\rangle_N
 \cr  & &\qquad\qquad
    +{1\over3}(4m_u-2m_d+m_s)\langle\bar u\gamma_0 u\rangle_N
    -{2\over3}(m_s+m_d+m_u)\langle\bar s\gamma_0 s\rangle_N
                       \Big\}\Big]\cr
 & &\qquad
    +{\M\over\MBS}\Big[\pi^4\Big\{ 
    {16\over9}\langle\bar dd\bar ss\rangle_N
    +{16\over9}\langle\bar uu\bar ss\rangle_N
    -{8\over9}\langle\bar dd\bar uu\rangle_N \cr
  & &\qquad\qquad  
    +{16\over9}\langle\bar dd\rangle_0
     (\langle\bar u\gamma_0 u\rangle_N+\langle\bar s\gamma_0 s\rangle_N)
    +{16\over9}\langle\bar uu\rangle_0
     (\langle\bar d\gamma_0 d\rangle_N+\langle\bar s\gamma_0 s\rangle_N) \cr
  & &\qquad\qquad -{8\over9}\langle\bar ss\rangle_0
     (\langle\bar d\gamma_0 d\rangle_N+\langle\bar u\gamma_0 u\rangle_N)
                       \Big\} \Big]
\end{eqnarray}
for the $\Lambda N$ channel and
\begin{eqnarray}\label{SR^indep}
&&-{1\over\MBS}\exp\left(-{\M^2\over\MBS}\right)\lambda^2\V^{indep.}\cr
 &=&{1\over4\pi^4}\Big\{
    \left(C_2\M\MB -C_3\MBS\right)
     \Big[\pi^2\Big\{-3\langle\d^\dagger\d\rangle_N
      -\langle\u^\dagger\u\rangle_N \Big\}
       -\pi^2\langle\bar\uu\rangle_N \Big]\cr
 & &\qquad
    +\left(C_1\M-C_2\MB\right)
     \Big[\pi^2\Big\{ 
     {1\over8}\left\langle{\alpha_s\over\pi}G^{\alpha\beta}G_{\alpha\beta}
             \right\rangle_N
    -{1\over6}\left\langle{\alpha_s\over\pi}{\cal S} 
                  [G^{\rho}_{0}G_{\rho0}]\right\rangle_N
\cr & &\qquad\qquad\qquad\qquad\qquad
    +{16\over3}i\langle{\cal S}[\bar\d\gamma_0 D_0\d]\rangle_N
    +{4\over3}i\langle{\cal S}[\bar\u\gamma_0 D_0\u]\rangle_N
\cr & &\qquad\qquad\qquad\qquad\qquad
    +m_{\u}\langle\bar\uu\rangle_N+2m_{\u}(\langle\bar\u\gamma_\mu\u\rangle_N
    -\langle\bar\d\gamma_\mu\d\rangle_N)
\Big\}\Big]\cr
 & &\qquad
    +{\M\over\MBS}\Big[{8\over3}\pi^4\langle\bar\dd 
                                                    \bar\dd\rangle_N
    +{16\over3}\pi^4\langle\bar\uu 
                                  \d^\dagger\d\rangle_N\Big]
 \Big\}
\end{eqnarray}
for other hyperon-nucleon channels.
Similarly, the sum rule for the spin-dependent part is
\begin{eqnarray}\label{SR^dep_L}
&&-{1\over\MBS}\exp\left(-{\M^2\over\MBS}\right)\lambda^2\V^{dep.}\cr
 &=&{1\over4\pi^4}r_k\Big\{
    \left(C_2\M\MB-C_3\MBS\right)
   \Big[\pi^2
    \Big\{{1\over18}\langle\bar d\gamma_k\gamma_5 d\rangle_N
         +{1\over18}\langle\bar u\gamma_k\gamma_5 u\rangle_N
         +{11\over9}\langle\bar s\gamma_k\gamma_5 s\rangle_N \cr
  & &\qquad\qquad
         -{2\over9}\langle\bar d\gamma_5i\sigma_{k0}d\rangle_N
         -{2\over9}\langle\bar u\gamma_5i\sigma_{k0}u\rangle_N
         +{2\over18}\langle\bar s\gamma_5i\sigma_{k0}s\rangle_N
    \Big\}\Big]\cr
 & &\qquad
     +\left(C_1\M-C_2\MB\right)\cr
 & &\qquad\times
   \Big[\pi^2\Big\{
      {4\over9}\langle{\cal S}[\bar d\gamma_k\gamma_5 iD_0 d]\rangle_N
     +{4\over9}\langle{\cal S}[\bar u\gamma_k\gamma_5 iD_0 u]\rangle_N
     -{20\over9}\langle{\cal S}[\bar s\gamma_k\gamma_5 iD_0 s]\rangle_N \cr
  & &\qquad\qquad 
    -{1\over3}(m_d+m_u-2m_s)
      \left(\langle\bar d\gamma_5i\sigma_{k0}d\rangle_N
    +\langle\bar u\gamma_5i\sigma_{k0}u\rangle_N
    -2\langle\bar s\gamma_5i\sigma_{k0}s\rangle_N\right) \cr
  & &\qquad\qquad
        -{1\over9}(3m_s-6m_u-4m_d)\langle\bar d\gamma_k\gamma_5 d\rangle_N
        -{1\over9}(3m_s-6m_d-4m_u)\langle\bar u\gamma_k\gamma_5 u\rangle_N\cr
  & &\qquad\qquad
        +{2\over9}(3m_d+3m_u-m_s)\langle\bar s\gamma_k\gamma_5 s\rangle_N
                       \Big\}\Big]\cr
 & &\qquad
    +{\M\over\MBS}\Big[\pi^4\Big\{
    -{16\over9}\langle\bar ss\rangle_0(\langle\bar d\gamma_5i\sigma_{k0}d
                     \rangle_N+\langle\bar u\gamma_5i\sigma_{k0}u\rangle_N)
    -{16\over9}(\langle\bar dd\rangle_0+\langle\bar uu\rangle_0)
                 \langle\bar s\gamma_5i\sigma_{k0}s\rangle_N \cr
  & &\qquad\qquad 
    +{8\over9}
     (\langle\bar dd\rangle_0\langle\bar u\gamma_5i\sigma_{k0}u\rangle_N
     +\langle\bar uu\rangle_0\langle\bar d\gamma_5i\sigma_{k0}d\rangle_N) \cr
  & &\qquad\qquad
    -{16\over9}\langle\bar dd\rangle_0(\langle\bar u\gamma_k\gamma_5 u
                     \rangle_N+\langle\bar s\gamma_k\gamma_5 s\rangle_N)
    -{16\over9}\langle\bar uu\rangle_0(\langle\bar d\gamma_k\gamma_5 d
                     \rangle_N+\langle\bar s\gamma_k\gamma_5 s\rangle_N) \cr
  & &\qquad\qquad 
    +{8\over9}\langle\bar ss\rangle_0(\langle\bar d\gamma_k\gamma_5 d
                     \rangle_N+\langle\bar u\gamma_k\gamma_5 u\rangle_N)
                          \Big\} \Big]
\end{eqnarray}
for the $\Lambda N$ channel and
\begin{eqnarray}\label{SR^dep}
&&-{1\over\MBS}\exp\left(-{\M^2\over\MBS}\right)\lambda^2\V^{dep.}\cr
 &=&{1\over4\pi^4}r_k\Big\{
    \left(C_2\M\MB-C_3\MBS\right)
    \Big[\pi^2\Big\{{5\over3}\langle\bar\d\gamma_k\gamma_5\d\rangle_N
               -{1\over3}\langle\bar\u\gvec\gamma_5\u\rangle_N\Big\}
     -i{1\over3}\pi^2\langle\bar\u\gamma_5\sigma_{k0}\u\rangle_N
 \Big]\cr
 & &\qquad
     +\left(C_1\M-C_2\MB\right)
    \Big[\pi^2\Big\{-{8\over3}\langle\bar\d\gamma_k\gamma_5 iD_0\d\rangle_N
         +{4\over3}\langle\bar\u\gamma_k\gamma_5 iD_0\u\rangle_N
\cr& &\qquad\qquad\qquad\qquad\qquad
         +{2\over3}m_{\u}(\langle\bar\u\gamma_\nu\gamma_5 u\rangle_N
           +3\langle\bar\d\gamma_\nu\gamma_5\d\rangle_N)\Big\}
\Big]\cr
 & &\qquad
    +{\M\over\MBS}\Big[
    -{16\over3}\pi^4\langle\bar\dd 
                                   \bar\d\gamma_5i\sigma_{k0}\d\rangle_N
    -{16\over3}\pi^4\langle\bar\uu 
                                   \bar\d\gamma_k\gamma_5\d\rangle_N
 \Big] \Big\}
\end{eqnarray}
for other hyperon-nucleon channels.
$C_1$, $C_2$ and $C_3$ are given by
\begin{eqnarray*}
  C_1
  &=&1-{1\over2}\left[\exp\left(-{\wp^2\over\MBS}\right)
                     +\exp\left(-{\wm^2\over\MBS}\right)\right],\cr
  C_2
  &=&-{1\over2}\left[{\wp\over\MB}\exp\left(-{\wp^2\over\MBS}\right)
               -{\wm\over\MB}\exp\left(-{\wm^2\over\MBS}\right)\right]
     +{\sqrt{\pi}\over4}\left[\Phi\left({\wp\over\MB}\right)
                             -\Phi\left({\wm\over\MB}\right)\right],\cr
  C_3
  &=&1-{1\over2}\left[
     \left(1+{\wp^2\over\MBS}\right)\exp\left(-{\wp^2\over\MBS}\right)
    +\left(1+{\wm^2\over\MBS}\right)\exp\left(-{\wm^2\over\MBS}\right)\right] ,
\end{eqnarray*}
where $\Phi$ is the error function.

From the vacuum-to-vacuum correlation function we obtain the Borel sum rule for the normalization constant $\lambda^2$ as~\cite{rf:RRY,rf:Ioffe}
\begin{eqnarray}\label{vSR_L}
 & &\lambda^2 {1\over\MBS}\exp\left(-{\M^2\over\MBS}\right)\cr
 &=&{1\over4\pi^4}\Big[
  - D_2\MBQ{1\over8}
  - D_1\Big\{
    {\pi^2\over8}
    \left\langle{\alpha_s\over\pi}G^{\alpha\beta}G_{\alpha\beta}\right\rangle_0
    +{1\over3}(3m_d+m_u-2m_s)\langle\bar dd\rangle
\cr& &\qquad\qquad
    +{1\over3}(3m_u+m_d-2m_s)\langle\bar uu\rangle
    +{1\over3}(3m_s-2m_d-2m_u)\langle\bar ss\rangle  \Big\}
\cr& &\qquad-{1\over\MBS}\pi^4\Big\{ {16\over9}\langle\bar dd\bar ss\rangle
                           +{16\over9}\langle\bar uu\bar ss\rangle
                           -{8\over9}\langle\bar dd\bar uu\rangle \Big\}\Big]
\end{eqnarray}
for the $\Lambda$ and
\begin{eqnarray}\label{vSR}
 & &\lambda^2 {1\over\MBS}\exp\left(-{\M^2\over\MBS}\right)\cr
 &=&{1\over4\pi^4}\Big[
  - D_2\MBQ{1\over8}
  - D_1\Big\{
    {\pi^2\over8}
    \left\langle{\alpha_s\over\pi}G^{\alpha\beta}G_{\alpha\beta}\right\rangle_0
    +m_{\u}\langle\bar\uu\rangle_0  \Big\}
  - {1\over\MBS}\Big\{\pi^4{8\over3}\langle\bar\dd\bar\dd\rangle_0\Big\}\Big]
\end{eqnarray}
for other hyperons. $D_1$ and $D_2$ are expressed by the effective threshold, $\w$, as
\begin{eqnarray*}
  D_1&=&1-\exp\left(-{\w^2\over\MBS}\right),\cr
  D_2&=&1-\left(1+{\w^2\over\MBS}+{1\over2}{\w^4\over\MBQ}\right)
              \exp\left(-{\w^2\over\MBS}\right).
\end{eqnarray*}
Dividing \Eq.{SR^indep_L} (\Eq.{SR^dep_L}) and \Eq.{SR^indep} (\Eq.{SR^dep}) by \Eq.{vSR_L} and \Eq.{vSR}, respectively, we finally obtain the sum rules for the spin-independent (spin-dependent) part of the hyperon-nucleon interaction strengths.

It is instructive to look at sum rules in the leading order of the OPE
before discussing the full results:
\begin{eqnarray*}
\V_{\Lambda p}
&=&{8\pi^2\over\MBS}\Big\{
{7\over6}\langle d^\dagger d\rangle_p+{7\over6}\langle u^\dagger u\rangle_p+{5\over3}\langle s^\dagger s\rangle_p
+{2\over3}\langle\bar dd\rangle_p+{2\over3}\langle\bar uu\rangle_p-{1\over3}\langle\bar ss\rangle_p\cr
& &\qquad
+\vec s\cdot\vec s\,'\,\Big[{1\over18}(-\Delta d-\Delta u-22\Delta s)
+{1\over9}(2\delta d+2\delta u-\delta s)\Big]\Big\}\cr
&=&{8\pi^2\over\MBS}\Big\{ 7.55 +\vec s\cdot\vec s\,'\,\Big[ 0.257
\Big]\Big\},\cr\cr
\V_{\Sigma^+ p}
&=&{8\pi^2\over\MBS}\Big\{
 3\langle u^\dagger u\rangle_p+\langle s^\dagger s\rangle_p
+\langle\bar ss\rangle_p
+\vec s\cdot\vec s\,'\,\Big[{1\over3}(-5\Delta u+\Delta s)
+{1\over3}\delta s\Big]\Big\}\cr
&=&{8\pi^2\over\MBS}\Big\{ 6.70 +\vec s\cdot\vec s\,'\,\Big[ -1.12
\Big]\Big\},\cr\cr
\V_{\Sigma^- p}
&=&{8\pi^2\over\MBS}\Big\{
 3\langle d^\dagger d\rangle_p+\langle s^\dagger s\rangle_p
+\langle\bar ss\rangle_p
+\vec s\cdot\vec s\,'\,\Big[{1\over3}(-5\Delta d+\Delta s)
+{1\over3}\delta s\Big]\Big\}\cr
&=&{8\pi^2\over\MBS}\Big\{ 3.70 +\vec s\cdot\vec s\,'\,\Big[ 0.527
\Big]\Big\},\cr\cr
\V_{\Xi^0 p}
&=&{8\pi^2\over\MBS}\Big\{
 3\langle s^\dagger s\rangle_p+\langle u^\dagger u\rangle_p
+\langle\bar uu\rangle_p
+\vec s\cdot\vec s\,'\,\Big[{1\over3}(-5\Delta s+\Delta u)
+{1\over3}\delta u\Big]\Big\}\cr
&=&{8\pi^2\over\MBS}\Big\{ 5.46 +\vec s\cdot\vec s\,'\,\Big[ 0.674
\Big]\Big\},\cr\cr
\V_{\Xi^- p}
&=&{8\pi^2\over\MBS}\Big\{
 3\langle s^\dagger s\rangle_p+\langle d^\dagger d\rangle_p
+\langle\bar dd\rangle_p
+\vec s\cdot\vec s\,'\,\Big[{1\over3}(-5\Delta s+\Delta d)
+{1\over3}\delta d\Big]\Big\}\cr
&=&{8\pi^2\over\MBS}\Big\{ 3.96 +\vec s\cdot\vec s\,'\,\Big[ -0.011
\Big]\Big\},
\end{eqnarray*}
where $\vec s$ ($\vec s\,'$) denotes the hyperon (nucleon) spin operator. 

For comparison, the interaction strength in the neutron-proton ($np$) channel is given by
\begin{eqnarray*}
\V_{n p}
&=&{8\pi^2\over\MBS}\Big\{
 3\langle d^\dagger d\rangle_p+\langle u^\dagger u\rangle_p
+\langle\bar uu\rangle_p
+\vec s\cdot\vec s\,'\,\Big[{1\over3}(-5\Delta d+\Delta u)
+{1\over3}\delta u\Big]\Big\}\cr
&=&{8\pi^2\over\MBS}\Big\{ 8.46 +\vec s\cdot\vec s\,'\,\Big[ 1.07
\Big]\Big\}.
\end{eqnarray*}

The spin-independent interaction strength is related to the vector and scalar charges of the nucleon while the spin-dependent strength to the axial and tensor charges.
In the $NN$ channel only up and down quark operators contribute, while in the $YN$ channels strange quark operators appear in addition.
It is interesting that the ratio of different flavor contributions is the same for the scalar and tensor charges in all $YN$ and $NN$ channels.
If the proton matrix elements of the up, down and strange quark operators are the same, all $YN$ interaction strengths coincide, which holds even if higher order terms of the OPE are included.
It should be noted, however, that this is not the $SU(3)$ limit. 
Even in the SU(3) limit the proton matrix elements of quark operators with different flavors are different in general.

Inserting the numerical values for the matrix elements we find that the spin-dependent interaction strength is considerably smaller than the spin-dependent interaction strength in each channel.
This is because the absolute values of the axial and tensor charges are considerably smaller than the vector and scalar charges except for the strange vector charge, which identically vanishes.
Also found is that both the spin-independent and spin-dependent interaction strengths are smaller in the $YN$ channel than in the $NN$ channel, since the strange charges are much smaller than the up and down charges.
An exception for this is the spin-dependent interaction strength in the $\Sigma^+p$ channel, which receives large contribution from the up axial charge.
Among $YN$ channels the $\Lambda N$ has the largest interaction strength due to the largest scalar charge contribution, which is even larger than in the $NN$ channel since the sign of the Wilson coefficients is oposite to the others.
If one looks at the charge dependence, the interaction strength is greater in the $\Sigma^+N$ ($\Xi^0N$) channel than in the $\Sigma^-N$ ($\Xi^-N$), since the up charges are larger than the down charges.

Let us now discuss the results including all the operators up to dimension 4 and the dimension 6 four-quark operators.
Up to now we have not explained how to determine $\w$, $\wp$ and $\wm$.
They are determined by the Borel stability analysis, i.e. , so that the calculated interaction strength has the most stable plateau as a function of the Borel mass squared.
We constrain $\wm$ to be greater than $\w$ since we expect the projection reduces the continuum contribution in the negative low energy region.
Also, we take $\wp$ to be equal to $\w$ for simplicity.
The interaction strength, however, is not sensitive to the choice of $\wp$.
Even if we take $|\w-\wp|=0.5$~GeV, the interaction strength changes only less than 3\%.
Fig.~1 shows the calculated interaction strength as a function of the Borel mass.
The upper and lower curves beyond $\MBS=1.2\;{\rm GeV}^2$ correspond to the spin-independent and spin-dependent interaction strengths, respectively.
There appears a stale plateau above $\MBS=1.4\;{\rm GeV}^2$ in all channels.
The prediction of the Borel sum rule is determined by taking the value in the region of the plateau.
The optimal value of the continuum threshold, $\wp$, in the $\Sigma^- p$ channel is much different from those in other channels.
However, even if we take $\wp$ in the $\Sigma^+ p$ channel to be the same as in the $\Sigma^- p$ channel, the difference in the interaction strength is less than 10\% for $\MBS=1.0\sim2.0\;{\rm GeV}^2$.
Fig.~2 shows the calculated interaction strengths without including the continuum.
From Fig.~1 and Fig.~2 we find that the contribution of the continuum is rather small.

Comparing the results in the leading order of the OPE and those including the higher order terms we find that the contribution of the higher order terms makes the strength decrease in the $\Sigma N$ channel but increase in the $\Xi N$ channel, relative to the strength in the $\Lambda N$ channel, so that the $\Sigma^+ p$ strength becomes smaller than the $\Xi^0 p$ strength.
We see that the spin-independent interaction strength in the $NN$ channel is greater than those in the $YN$ channels, which is similar to the results in the leading order.
We should, however, keep in mind uncertainties in the matrix elements and the continuum contributions and therefore conclude at the moment that the strengths in the $np$, $\Lambda p$ and $\Xi^0p$ channels are similar.
The strengths are smaller in the $\Sigma p$ and $\Xi^-p$ channels than in other three channels.
Furthermore, the $\Sigma^-p$ interaction is the weakest in all the hyperon-proton interactions.

Comparing the spin-independent part with the spin-dependent part we find that the spin-dependent interaction is much smaller than the spin-independent interaction in each channel.
The interaction strengths in the $\Sigma^+ p$ and $\Xi^-p$ channels have an opposite sign to the interaction strengths in other channels. 
The spin-singlet interaction is greater than the spin-triplet one in the $\Sigma^+ p$ and $\Xi^-p$ channels but the relation is opposite in other channels.

Up to now we have discussed the results of sum rules identifying them as interaction strengths because interaction strengths are convenient for comparing different hyperon-nucleon and nucleon-nucleon channels.
Now, we want to compare our results with those of phenomenological meson-exchange models.
For this purpose, scattering lengths are more convenient.
Therefore, from now on we identify our results as scattering lengths ignoring $O(a^2)$ differences.
Table~I shows calculated scattering lengths with estimated errors for the hyperon-nucleon channels in fm.
The errors due to uncertainties of the spin-dependent nucleon matrix elements are as small as $\pm0.04\sim\pm0.07$~fm. 
Among the spin-independent nucleon matrix elements the scalar charges cause most of errors. 
The uncertainties of $\langle\bar u u\rangle_p$ and $\langle\bar d d\rangle_p$ are about 30\% and that of $\langle\bar s s\rangle_p$ is about 60\%.
These errors change the scattering lengths by $\pm0.4\sim\pm1.2$~fm.
$\langle{\cal S}[\bar q\gamma_\mu D_\nu q]\rangle_p$ is expected to have much smaller error than $\langle\bar qq\rangle_p$~\cite{rf:GRV}.
The uncertainty of $\langle{\alpha_s\over\pi}G_{\mu\nu}G^{\mu\nu}\rangle_p$ is about 20\%, which changes $a_{\Sigma p}$ by only about 2\% and affects others less than 0.3\%.
The errors of other spin-independent nucleon matrix elements are also thought to be small since they are higher order terms.
Furthermore, the vacuum condensates of the quark and the gluon operators have uncertainties about 20\% and the strange quark mass lies in the range between 100~MeV and 200~MeV.
The total error due to these uncertainties is about 15\%.
The errors of the scattering lenghts due to the uncertainties of all these input parameters are shown in Table~I.
Even though we did not include in the errors of Table~I, we also estimated errors due to effective continuum thresholds which change $a_{YN}$ roughly by $\pm1$~fm.

Table~II shows the scattering lengths calculated in the Nijmegen meson-exchange models, NSC97a-f~\cite{Nmodel}.
Comparing Table~I and Table~II, we find that the scattering lengths obtained by sum rules are rather larger than those in the meson-exchange models.
It should be remembered, however, that the results of sum rules coincide with the scattering lengths only when the scattering lengths are small.
The results of the sum rules and the meson-exchange models have some common tendencies: the $\Sigma^+p$ channel has stronger spin dependence than the $\Lambda p$ channel and the triplet scattering length is smaller than the singlet scattering length in the ${\Sigma^+p}$ channel.
On the contrary, the results are quit different from each other in the $\Xi^0p$ channel.
This might be related to anormalous behavior of the effective range in the Nijmegen models, which is extremely large in the triplet channel and is negative and large in the singlet channel.

A comment on the spin dependence of the interaction in the $\Lambda N$ channel and observed hypernuclear states is in order here.
Experimentally, the ground state of ${}^4_{\Lambda}{\rm He}$-hypernucleus is known to be $0^+$ and $1^+$ is an exited state.
This seems to contradict with the present result that the triplet $\Lambda p$ interaction is more attractive than the singlet one.
According to Akaishi et al., however, if one takes the three-body force into account the ordering of the observed ${}^4_{\Lambda}{\rm He}$-hypernuclear states is reproduced with the Nijmegen model D~\cite{rf:Akaishi} in which the triplet scattering length is a little greater than the singlet one\cite{rf:Dmodel}.
Even though this is not conclusive, this at least tells us that one must be careful when one extracts information on the hyperon-nucleon interaction from observed hypernuclear states.

\section{Summary}
In summary, we have formulated the QCD sum rule for the $YN$ interactions starting from the spin-dependent hyperon correlation function, whose matrix element is taken with respect to the one-nucleon state.
The dispersion integral around the threshold is identified as a measure of the interaction strength in the $YN$ channel.
The Wilson coefficients of the OPE for the hyperon correlation functions have been calculated.
Strange quark operators appear as well as up and down quark operators.
After Borel transforming we have obtained Borel sum rules which relate the $YN$ interaction strengths to the nucleon matrix elements of quark-gluon operators.
We have then discussed physical implications of the obtained sum rules for the interaction strengths in the $YN$ and the $NN$ channels.
It was found that the interaction strength is smaller in the $YN$ channels than in the $NN$ channel since the matrix elements of strange quark operators are smaller than those of up and down quark operators.
Among $YN$ channels the $\Lambda N$ channel turned out to have a stronger interaction strength than the $\Sigma N$ and $\Xi N$ channels.
It was also found that the $\Sigma^+N$ ($\Xi^0N$) interaction strength is greater than the $\Sigma^-N$ ($\Xi^-N$) interaction strength since matrix elements of up quark operators are greater than those of the down quark operators.
Taking uncertainties into account, however, we tentatively concluded that the $np$, $\Lambda p$ and $\Xi^0p$ channels have similar interaction strengths.
It turned out that the strengths are smaller in the $\Sigma p$ and $\Xi^-p$ channels than in other channels and also that the $\Sigma^-p$ channel has the weakest interaction in all channels.
Also found is that the spin-dependent strength is smaller than the spin-independent strength in all channels.
The spin-dependence in the $\Sigma^+ p$ and $\Xi^-p$ channels is oposite to other channels.
The results are compared with those of the phenomenological meson-exchange models.
The results of the sum rules and the meson-exchange models have some common tendencies.
They are, however, different in some other respects. 

In the present paper we have ignored the coupling of different hyperon-nucleon and hyperon-hyperon channels such as $\Lambda N$-$\Sigma N$ and $\Xi N$-$\Sigma\Sigma$-$\Lambda\Lambda$.
One might be able to formulate the sum rule for the transition, $\Lambda N\rightarrow\Sigma N$, by considering the time-ordered product of the sigma field and the lambda field, which is an interesting extension of the present approach.
Another interesting possibility is to study hyperon matrix elements of the quark-gluon composite operators by means of the sum rule.
This would be done by constructing sum rule starting from the hyperon matrix element of the nucleon correlation function and relating the obtained hyperon-nucleon interaction strengths with the ones in the present paper.
These applications would provide us with important new information about hyperon-nucleon interactions and hyperon structures.

\acknowledgments

We would like to thank Prof. Yoshinori Akaishi for valuable comments.


\begin{table}[t]
\caption{The calculated scattering lengths for the hyperon-nucleon channels (uint in fm).}
\vspace*{10pt}
{\setlength{\tabcolsep}{10pt}
\begin{tabular}{c c c}
Channel& spin-triplet&spin-singlet \cr\hline
$\Lambda p$  & $5.7\pm1.0$ & $4.8\pm1.0$ \cr
$\Sigma^+p$  & $3.5\pm0.8$ & $5.6\pm0.8$ \cr
$\Sigma^-p$  & $2.1\pm0.6$ & $1.1\pm0.6$ \cr
$\Xi^0p$     & $6.0\pm1.4$ & $3.4\pm1.4$ \cr
$\Xi^-p$     & $3.5\pm1.1$ & $3.8\pm1.1$ \cr
\end{tabular}
}
\end{table}

\begin{table}
\caption{The triplet $^3S_1$ and singlet $^1S_0$ scattering lengths in fm for models NSC97a-f in the different channels.}
\vspace*{10pt}
{\setlength{\tabcolsep}{10pt}
\begin{tabular}{c c c c c c c}
        &\multicolumn{2}{c}{$\Lambda p$}
        &\multicolumn{2}{c}{$\Sigma^+p$}
        &\multicolumn{2}{c}{$\Xi^0p$} \cr
 Model  &$^3S_1$&$^1S_0$ &$^3S_1$&$^1S_0$ &$^3S_1$&$^1S_0$ \cr\hline
(a)  & $2.18$ & $0.71$  & $0.14$ & $4.35$  & $0.038$ & $-0.46$ \cr
(b)  & $2.13$ & $0.90$  & $0.17$ & $4.32$  & $0.045$ & $-0.45$ \cr
(c)  & $2.08$ & $1.20$  & $0.25$ & $4.28$  &$-0.001$ & $-0.43$ \cr
(d)  & $1.95$ & $1.71$  & $0.29$ & $4.23$  &$-0.041$ & $-0.42$ \cr
(e)  & $1.86$ & $2.10$  & $0.28$ & $4.23$  &$-0.050$ & $-0.41$ \cr
(f)  & $1.75$ & $2.51$  & $0.25$ & $4.35$  & $0.030$ & $-0.40$ \cr
\end{tabular}
}
\end{table}

\begin{figure}
  \begin{center}
    \leavevmode
  \epsfxsize=6.0 in
  \epsfbox{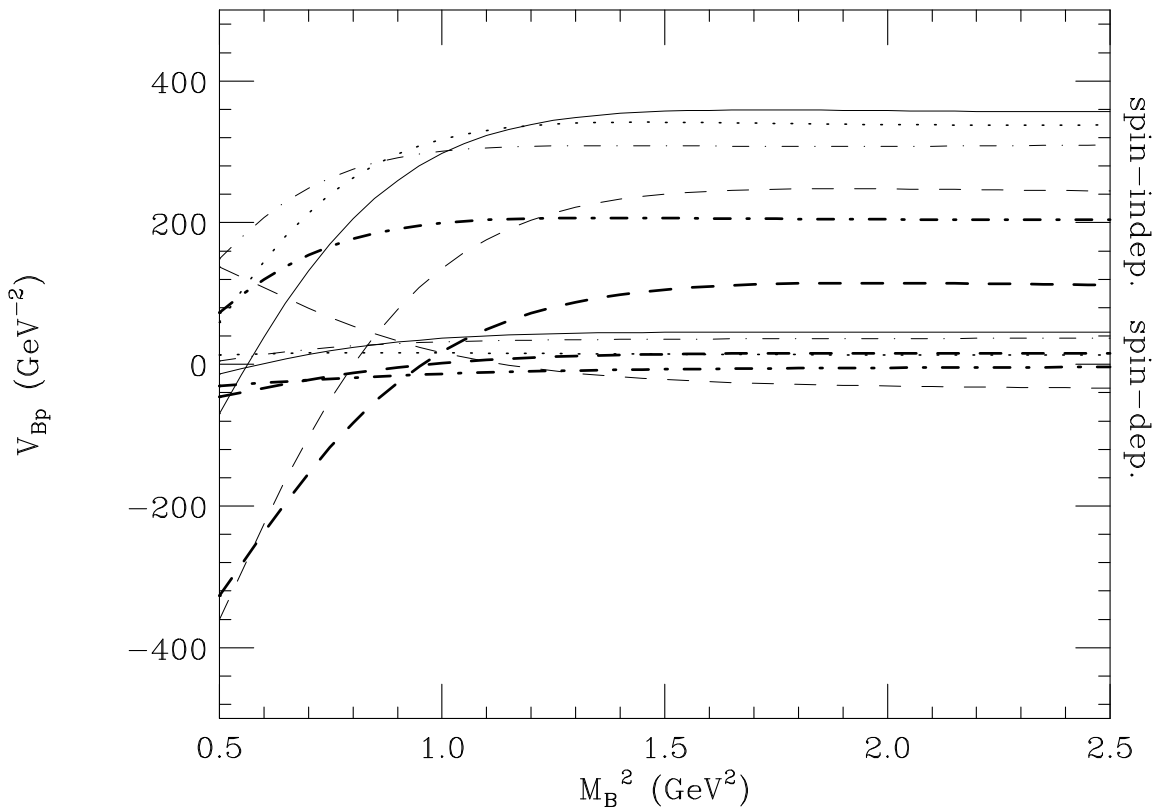}
  \end{center}
\caption{
The interaction strength, $v_{B p}$,
 as a function of the Borel mass where $\omega_0=2.8$ GeV and $\omega_-=3.4$ GeV for the  $np$ channel, $\omega_0=2.7$ GeV  and $\omega_-=3.3$ GeV for the $\Lambda p$ channel, $\omega_0=3.1$ GeV  and $\omega_-=4.0$ GeV for the $\Sigma^+ p$ channel, $\omega_0=3.5$ GeV  and $\omega_-=10.8$ GeV for the $\Sigma^- p$ channel, $\omega_0=2.4$ GeV  and $\omega_-=2.8$ GeV for the $\Xi^- p$ channel and $\omega_0=2.5$ GeV  and $\omega_-=2.9$ GeV for the $\Xi^- p$ channel.
The $np$, $\Lambda p$, $\Sigma^+ p$, $\Sigma^- p$, $\Xi^+ p$ and $\Xi^- p$ channels are shown as solid, dotted, dashed, thick dashed, dot-dashed and thick dot-dashed lines, respectively. 
}
  \label{fig1}
\end{figure}

\begin{figure}
  \begin{center}
    \leavevmode
  \epsfxsize=6.0 in
  \epsfbox{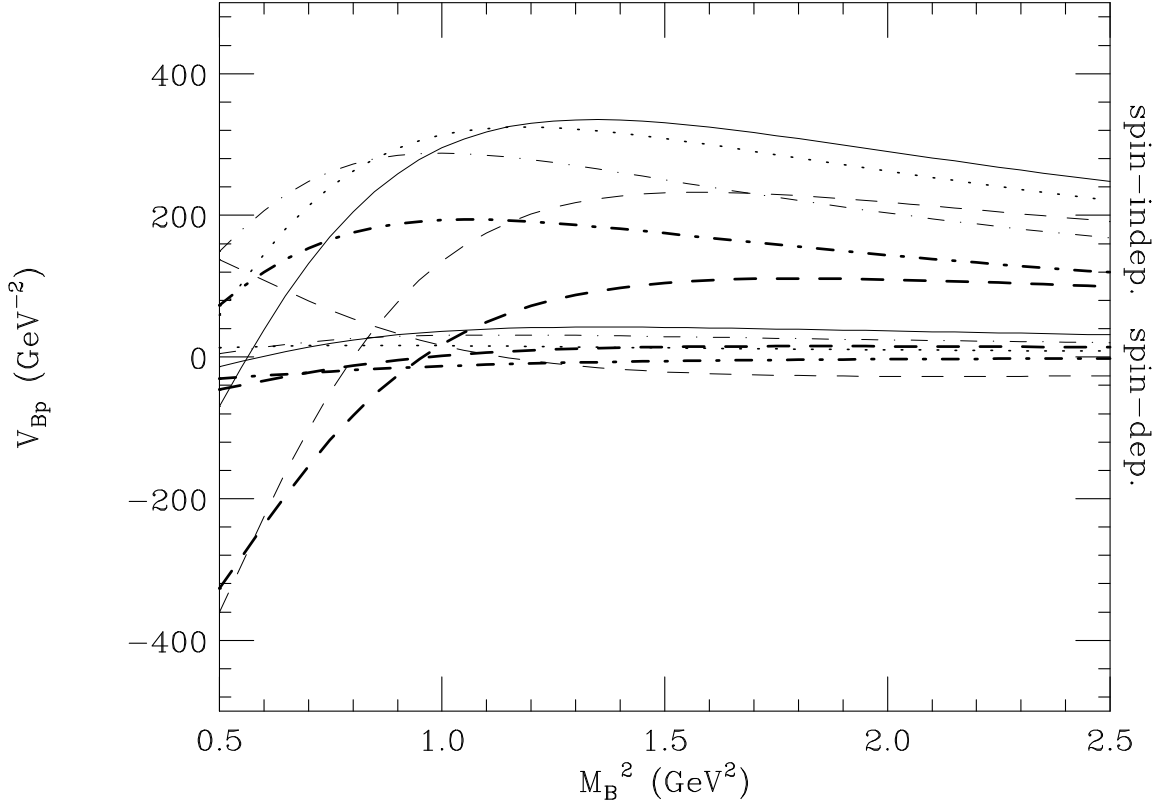}
  \end{center}
\caption{
The interaction strength, $v_{B p}$,
 as a function of the Borel mass where $\omega_0=\infty$ and $\omega_\pm=\infty$.
The $np$, $\Lambda p$, $\Sigma^+ p$, $\Sigma^- p$, $\Xi^+ p$ and $\Xi^- p$ channels are shown as solid, dotted, dashed, thick dashed, dot-dashed and thick dot-dashed lines, respectively. 
}
  \label{fig2}
\end{figure}

\end{document}